\documentclass[twocolumn,prd,superscriptaddress,altaffilletter,nofootinbib]{revtex4-1}
\usepackage{float}
\usepackage{amsmath}
\usepackage{amssymb}
\usepackage{amsfonts}
\usepackage{comment}
\usepackage{graphicx}
\usepackage{hyperref}
\usepackage{interval}
\usepackage[nolist]{acronym}
\usepackage[usenames,dvipsnames]{xcolor}
\usepackage{xspace}
\usepackage{subfigure}
\usepackage{tikz}
\usepackage[squaren]{SIunits}
\usetikzlibrary{arrows,shapes,trees,decorations.pathreplacing}
\allowdisplaybreaks

\newcommand{\likelihood}{\mathcal{L}}

\newcommand{\Dh}{\vec{D_H}}
\newcommand{\Ob}{\vec{O}}
\newcommand{\SNR}{\vec{\rho}}
\newcommand{\CHISQ}{\vec{\xi^2}}

\newcommand{\DPHI}{\Delta \vec{\phi}}
\newcommand{\DT}{\Delta \vec{t}}
\newcommand{\idq}{\log \likelihood_{\text{iDQ}}}
\newcommand{\gidq}{\log \widehat{\likelihood}_{\text{iDQ}}}
\newcommand{\cprob}[2]{P (\, #1 \mid #2 \,) } 

\begin{document}

\title{Incorporation of Statistical Data Quality Information into the GstLAL Search Analysis}

\author{Patrick Godwin}
\affiliation{Department of Physics, The Pennsylvania State University, University Park, PA 16802, USA}
\affiliation{Institute for Gravitation and the Cosmos, The Pennsylvania State University, University Park, PA 16802, USA}

\author{Reed Essick}
\affiliation{Kavli Institute for Cosmological Physics, The University of Chicago, Chicago, Illinois, 60637, USA}

\author{Chad Hanna}
\affiliation{Department of Physics, The Pennsylvania State University, University Park, PA 16802, USA}
\affiliation{Institute for Gravitation and the Cosmos, The Pennsylvania State University, University Park, PA 16802, USA}
\affiliation{Department of Astronomy and Astrophysics, The Pennsylvania State University, University Park, PA 16802, USA}
\affiliation{Institute for CyberScience, The Pennsylvania State University, University Park, PA 16802, USA}

\author{Kipp Cannon}
\affiliation{RESCEU, The University of Tokyo, Tokyo, 113-0033, Japan}

\author{Sarah Caudill}
\affiliation{Nikhef, Science Park, 1098 XG Amsterdam, Netherlands}

\author{Chiwai Chan}
\affiliation{RESCEU, The University of Tokyo, Tokyo, 113-0033, Japan}

\author{Jolien D. E. Creighton}
\affiliation{Leonard E.\ Parker Center for Gravitation, Cosmology, and Astrophysics, University of Wisconsin-Milwaukee, Milwaukee, WI 53201, USA}

\author{Heather Fong}
\affiliation{RESCEU, The University of Tokyo, Tokyo, 113-0033, Japan}
\affiliation{Graduate School of Science, The University of Tokyo, Tokyo 113-0033, Japan}

\author{Erik Katsavounidis}
\affiliation{LIGO, Massachusetts Institute of Technology, Cambridge, MA 02139, USA}

\author{Ryan Magee}
\affiliation{Department of Physics, The Pennsylvania State University, University Park, PA 16802, USA}
\affiliation{Institute for Gravitation and the Cosmos, The Pennsylvania State University, University Park, PA 16802, USA}

\author{Duncan Meacher}
\affiliation{Leonard E.\ Parker Center for Gravitation, Cosmology, and Astrophysics, University of Wisconsin-Milwaukee, Milwaukee, WI 53201, USA}

\author{Cody Messick}
\affiliation{Department of Physics, The Pennsylvania State University, University Park, PA 16802, USA}
\affiliation{Institute for Gravitation and the Cosmos, The Pennsylvania State University, University Park, PA 16802, USA}

\author{Soichiro Morisaki}
\affiliation{Institute for Cosmic Ray Research, The University of Tokyo, 5-1-5 Kashiwanoha, Kashiwa, Chiba 277-8582, Japan}

\author{Debnandini Mukherjee}
\affiliation{Department of Physics, The Pennsylvania State University, University Park, PA 16802, USA}
\affiliation{Institute for Gravitation and the Cosmos, The Pennsylvania State University, University Park, PA 16802, USA}

\author{Hiroaki Ohta}
\affiliation{RESCEU, The University of Tokyo, Tokyo, 113-0033, Japan}

\author{Alexander Pace}
\affiliation{Department of Physics, The Pennsylvania State University, University Park, PA 16802, USA}
\affiliation{Institute for Gravitation and the Cosmos, The Pennsylvania State University, University Park, PA 16802, USA}

\author{Iris de Ruiter}
\affiliation{Astronomical Institute, Anton Pannekoek, University of Amsterdam, Science Park 904, 1098 XH Amsterdam, Netherlands}
\affiliation{Nikhef, Science Park, 1098 XG Amsterdam, Netherlands}

\author{Surabhi Sachdev}
\affiliation{Department of Physics, The Pennsylvania State University, University Park, PA 16802, USA}
\affiliation{Institute for Gravitation and the Cosmos, The Pennsylvania State University, University Park, PA 16802, USA}
\affiliation{LIGO Laboratory, California Institute of Technology, MS 100-36, Pasadena, California 91125, USA}

\author{Leo Tsukada}
\affiliation{RESCEU, The University of Tokyo, Tokyo, 113-0033, Japan}
\affiliation{Graduate School of Science, The University of Tokyo, Tokyo 113-0033, Japan}

\author{Takuya Tsutsui}
\affiliation{RESCEU, The University of Tokyo, Tokyo, 113-0033, Japan}

\author{Koh Ueno}
\affiliation{RESCEU, The University of Tokyo, Tokyo, 113-0033, Japan}

\author{Leslie Wade}
\affiliation{Department of Physics, Hayes Hall, Kenyon College, Gambier, Ohio 43022, USA}

\author{Madeline Wade}
\affiliation{Department of Physics, Hayes Hall, Kenyon College, Gambier, Ohio 43022, USA}

\date{\today}
\begin{abstract}
We present updates to GstLAL, a matched filter gravitational-wave search pipeline,
 in Advanced LIGO and Virgo's
third observing run. We discuss the incorporation of statistical data quality
information into GstLAL's multi-dimensional likelihood ratio ranking statistic and
additional improvements to search for gravitational wave candidates found in only
one detector.
Statistical data quality information is provided by iDQ, a data quality pipeline
that infers the presence of short-duration transient noise
in gravitational-wave data using the interferometer's auxiliary state, which has
operated in near real-time since before LIGO's first observing run in 2015.
We look at the performance and impact on noise rejection by the inclusion of
iDQ information in GstLAL's ranking statistic, and discuss
GstLAL results in the GWTC-2 catalog, focusing on two
case studies; GW190424A, a single-detector gravitational-wave event
found by GstLAL and a period of time in Livingston impacted by a thunderstorm.
\end{abstract}

\maketitle

\section{Introduction}\label{sec:intro}

During Advanced LIGO and Virgo's second observing run, the detection of a
binary neutron star merger, GW170817 \cite{TheLIGOScientific:2017qsa} and its joint
detection of a short gamma ray burst \cite{Monitor:2017mdv,Savchenko:2017ffs}
observed by Fermi-GBM and INTEGRAL sparked a massive follow-up campaign
with more than 70 telescopes and observatories participating \cite{GBM:2017lvd}.
GW170817 started a new era of multi-messenger astronomy and provided
insight into compact binary mergers and their associated
electromagnetic counterparts, prompting an avalanche of further study, from advancing
our understanding of short Gamma-ray burst beaming profiles
(e.g. \cite{Monitor:2017mdv, Mooley2018, Farah:2019tue}), kilonova light-curves and
evolution (e.g. \cite{Abbott_2017, Dietrich:2020lps, Coulter:2017wya, Kilpatrick:2017mhz}),
cosmology and the expansion rate of the universe \cite{Abbott:2019yzh, Fishbach:2018gjp},
the equation of state of dense nuclear matter
(e.g. \cite{PhysRevLett.121.161101, PhysRevD.101.063007}), and the mass distribution
of compact objects \cite{Abbott_2019} among countless others.

GW170817 was identified in low-latency as a single-detctor candidate in the LIGO
Hanford detector by GstLAL, a matched filter gravitational-wave search for compact
binary coalescences \cite{Messick:2016aqy, Sachdev:2019vvd}.
However, at that time both the LIGO Livingston and Virgo detectors also were recording
science-quality data.
LIGO Livingston, in particular, was expected to have witnessed a similar signal
to the Hanford detector, and it was quickly discovered that a non-Gaussian
noise transient, or \emph{glitch} \cite{TheLIGOScientific:2016zmo}, in the Livingston
interferometer coincided with GW170817's inspiral track, thereby causing GstLAL
to neglect data from that detector at that time.
While the noise transient was from a familiar class of such glitches and was later
modeled and subtracted from the detector data, enabling precision measurements
of the gravitational-wave signal using Livingston data
\cite{Pankow:2018qpo, PhysRevX.9.011001}, the presence of non-Gaussian noise
initially complicated the detection process, highlighting the need for rapid,
reliable data quality information.

iDQ, a statistical inference framework that generates probabilistic data quality
information in low-latency \cite{idq-methods}, has operated throughout the
advanced detector era.
iDQ autonomously identified auxiliary witnesses to the type of
non-Gaussian noise coincident with GW170817 and correctly labeled the subsecond
interval containing the artifact as extremely likely to contain a glitch in low-latency,
less than 8 seconds after GstLAL reported the candidate.
Unfortunately, at that time, iDQ's statistical data quality information was not
incorporated into the GstLAL search.
Nonetheless, iDQ's autonomous identification of the noise transient, along with
further human vetting, allowed GW170817 to be announced to the broader
astrophysical community in time to inform electromagnetic follow-up~\cite{GW170817GCN}.

GW170817 points out the benefits of folding probabilistic data quality information
into searches directly.
This work describes the incorporation of statistical data quality information
from iDQ within the GstLAL search during the third observing run, an important
milestone in mitigating the impact of non-Gaussian noise within searches in
low-latency.

In April 2019, the advanced LIGO \cite{TheLIGOScientific:2014jea} and Virgo
\cite{TheVirgo:2014hva} interferometers started their third
observational run, O3, and with it, began sending out automated open public alerts
from gravitational-wave candidates.
The goal of low-latency gravitational-wave detection
and alerting infrastructure \cite{LIGOScientific:2019gag} is to detect the onset
of electromagnetic emission coming from compact binary coalesences within
seconds of the merger \cite{Cannon:2011vi}.
This requires the need for rapid validation and follow-up of
gravitational-wave candidates.

Non-Gaussian noise in LIGO and Virgo can produce
short-duration noise transients in the gravitational-wave strain, $h(t)$,
as seen in GW170817.
The presence of glitches in gravitational-wave
data limits the search sensitivity to gravitational-wave signals.
Being able to identify periods of
non-Gaussianities and limit their effect on searches is essential for
detecting gravitational-wave signals.
The increased sensitivity of gravitational-wave
detectors in O3 has increased the rate at which gravitational-wave detections are
made. However, the rate of glitches has also increased in both LIGO detectors
during the first half of advanced LIGO and Virgo's third observing run, O3a \cite{gwtc-2},
making the use of automated noise mitigation methods and rapid follow-up more
critical.

The detector's auxiliary state in advanced LIGO \cite{TheLIGOScientific:2014jea}
is monitored by the
instrumental control system, which includes the optical configuration
\cite{Mueller:2016hex}, seismic
isolation \cite{Matichard:2015eva}, and many control systems which are
responsible for keeping the
interferometer in lock \cite{Staley:2015nie,Rollins:2016hlk}.
Beyond this, many sensors monitor the detector's
environment \cite{Effler:2014zpa}. In total, $O(10^5)$ channels monitor
the detector's auxiliary state in advanced LIGO.
The noise sources that produce glitches in $h(t)$ are occasionally
witnessed by at least one of the
detector's many subsystems or in its surrounding environment \cite{Effler:2014zpa},
which can be detected in one or more auxiliary channels. 

Understanding the physical mechanism in which glitches couple to $h(t)$ allows
one to either identify and remove the source of the noise or identify times
when glitches due to this coupling are present \cite{Nuttall:2015dqa}.
If a physical mechanism can not be observed, one has to rely on
statistical couplings to infer the presence of a glitch, by identifying statistical
correlations between $h(t)$ and auxliary channels. Indeed, given the large
number of auxiliary channels, the vast majority of couplings to $h(t)$ remain
unmeasured and statistical inference remains the best chance to systematically
identify terrestrial noise sources. iDQ provides a framework
to infer the presence of glitches in $h(t)$ by identifying statistical
correlations between $h(t)$ and auxiliary channels, as well as providing a mechanism
to discover new statistical couplings as they arise in low-latency.
It has operated in near-real-time since before O1 in 2015, and was sped up for
the latest observing run to produce
probabilistic data quality information encoded as a set of streaming timeseries
available concurrently with $h(t)$ in low-latency.

Historically, gravitational-wave searches handle times associated with poor
data quality by removing the problematic data from being analyzed or by
vetoing any candidates during this time \cite{TheLIGOScientific:2017lwt}.
These periods of poor data quality
are flagged by various data quality vetoes based on their severity
\cite{LIGOScientific:2019hgc,TheLIGOScientific:2016zmo,Aasi:2014mqd},
and are either generated in low-latency from
select known couplings or produced in an offline fashion after various
data quality investigations. One benefit of using non-binary data products
is the ability to continuously downrank periods associated with various
non-Gaussianities. This allows searches to detect extremely confident astrophysical
signals in the presence of excess non-Gaussian noise instead of forcing them
to reject the entire stretch of data, regardless of whether there is a signal
present, based on non-Gaussian noise.
In this manner, iDQ
can be used to fold in data quality information into a gravitational-wave
search without vetoing an astrophysical event coincident with a glitch,
as in the case with GW190424A, discussed in Sec. \ref{sec-GW190424A}.

GstLAL is a matched filter gravitational-wave
search pipeline aimed at detecting compact binary coalesences in near real-time,
providing event significance estimates and point estimates for binary parameters.
GstLAL operates in both low-latency for near real-time detection, as well as an offline
mode to process gravitational-wave data with background statistics collected over
the analysis period as well as additional information generated offline such as
data quality vetoes and improvements in $h(t)$ calibration \cite{Cahillane:2017vkb}.
GstLAL identifies candidates by using a multi-dimensional likelihood ratio ranking
statistic $\likelihood$, described in 
\cite{Messick:2016aqy,Sachdev:2019vvd,Hanna:2019ezx,Cannon:2015gha,Cannon:2008zz},
which folds in information about the
candidate event's SNR in each detector, a multi-detector signal consistency test
\cite{Hanna:2019ezx}, time-averaged detector sensitivity, the signal population model, and
information about the collected background.

In this work, we describe methods of handling transient noise in GstLAL in Sec.
\ref{sec:gstlal-dq},
and discuss the inclusion of statistical data quality information via iDQ
into a matched filter search, by including it as a term in GstLAL's ranking
statistic in Sec. \ref{sec:ranking-stat}. We also discuss the performance
and pipeline sensitivity from
its inclusion and its impact on the GWTC-2 catalog in Sec. \ref{sec:results}.
In particular, we look at
two case studies; GW190424A, a single-detector gravitational-wave event
found by GstLAL which is found to be vetoed by a data quality flag and a period
of non-Gaussianity caused by a nearby thunderstorm in LIGO-Livingston and identified
by iDQ.

\section{Incorporation of Statistical Data Quality Information} \label{sec:gstlal-dq}

As mentioned in Sec. \ref{sec:intro}, gravitational-wave searches have
historically handled periods of poor data quality by either vetoing candidates
during this time at the event identification stage or by removing such segments
from $h(t)$ in the data conditioning stage, through a procedure called gating
\cite{Messick:2016aqy,Usman:2015kfa}.
GstLAL handles times associated with poor data quality with several complimentary
approaches; (1) signal-based vetoes using a $\xi^2$ test \cite{Messick:2016aqy},
(2) an auto-gating
procedure based on loud excursions from $h(t)$ \cite{Messick:2016aqy}, and (3),
gating times associated with poor data quality from a known list of data quality
vetoes \cite{Aasi:2014mqd}.

GstLAL filters
the gravitational-wave strain, $h(t)$, against a bank of known compact binary
template waveforms. Both the data and the template waveforms are whitened using
a measured noise power spectral density (PSD) estimate. The template waveforms
are whitened ahead of time using a reference PSD. The data is whitened on-the-fly
during operation to track drifts in the PSD over a detector lock segment, using
a median-mean PSD estimation approach \cite{Messick:2016aqy}. This provides an
estimation technique which is robust against short time-scale fluctuations,
including many glitch classes. 

Different categories of data quality vetoes are treated based on their severity.
Of interest in compact binary searches are
category 1 (CAT1) vetoes, flagging exceptionally egregious times, 
and category 2 (CAT2) vetoes based on physical couplings.
Times associated with CAT1 vetoes are not analyzed
to avoid causing issues with PSD estimation. Gating based on loud $h(t)$ excursions
and CAT2 data quality vetoes
are treated similarly; the target data associated with a segment is padded on both
sides and a windowing function is applied to it. During the first observational
run, O1, a square window was used, replacing the data within the window with zeros.
In the second observational run, O2, a Tukey window was used to smoothly transition
between the gate and the whitened data. In addition, a trigger from an individual
detector is vetoed if it coincides with a gate.

The procedure discussed in this paper replaces the treatment of vetoes in (3),
and instead relies
on iDQ to infer the presence of non-Gaussian noise in the data. Although Ref.
\cite{idq-methods} discusses many possible ways to do this, we focus on a relatively
simple, yet effective, scheme.
This is done
by adding a term in GstLAL's multi-dimensional ranking statistic that includes
information from iDQ to downrank candidate events that are in close proximity
with a glitch. This term is targetted at single detector candidates rather
than events found in coincidence, as discussed in \ref{sec:ranking-stat}, since
many of the consistency checks found for coincident events are not present
for single detector candidates.
Among other data products, iDQ produces a log-likelihood estimate
of non-Gaussian noise, $\idq$. We do not use this statistic as-is, however. Instead,
we renormalize $\idq$ to better control its possible impact on the GstLAL search.
This renormalized quantity, $\gidq$, is maximized over a $\pm 1$ second window around
the candidate event.

One benefit of this approach is that data quality information from external
sources can be added in without vetoing gravitational-wave events which
coincide with periods of poor data quality. Furthermore, iDQ data products
are available in low-latency at the same time as $h(t)$, so this work can be extended
to run in low-latency operation as well, as would be required for astrophysical events
coincidence with non-Gaussian noise like GW170817.

\subsection{iDQ Renormalization Procedure}


iDQ produces data products that are available at the same time as gravitational-wave
strain data in its low-latency operation, but due to its low-latency nature can be subject
to data
dropouts or sub-optimal calibration during periods of extreme non-Gaussianity.
In order to mitigate such effects from the low-latency runs, $\idq$ timeseries
were collected across the time of interest and renormalized. This also allows us
to better control the possible impact that iDQ information could have on the search.
Since the period
of interest is over a period of several months, we instead collect data over smaller periods
of time, about a week of coincident data, and renormalized for each stretch of time.

For each timespan, the $\idq$ timeseries
is ingested and the raw timeseries sampled at 128 Hz is aggregated and maximized over
1 second windows. These aggregated values are used to calculate
the transformation, defined as

\begin{equation}
\begin{split}
 \label{ch5-eq:transform}
 \gidq &=\begin{cases} 0 & P < P_{min} \\ \log {\frac{P}{100 - P}} & P_{min} \leq P \leq P_{max} \\ 15 & P \geq P_{max} \end{cases}, \\
 P_{min} &= 50, \\
 P_{max} &= \frac{100}{1+e^{-\max{(\gidq)}}} = \frac{100}{1+e^{-15}}.
\end{split}
\end{equation}
where $P$ is the percentile of $\gidq$ values over a given
chunk.

\begin{figure}[!ht]
    \centering
    \includegraphics[width=\columnwidth]{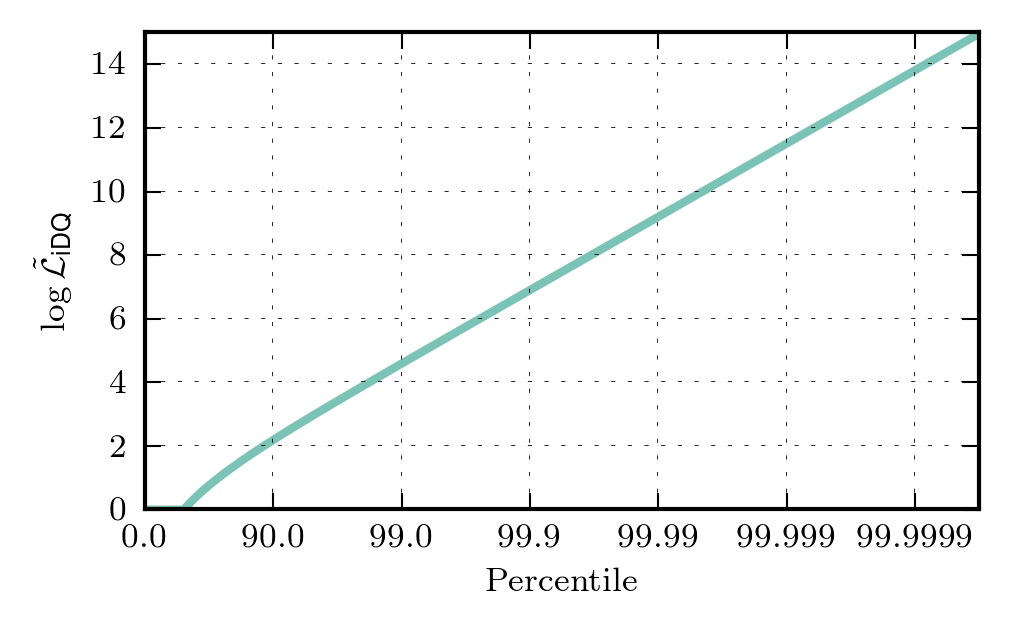}
    \caption{
        Mapping between $\idq$ and $\gidq$ based on percentile. The effects
        of limiting the dynamic range based on $P_{min}$ and $P_{max}$
        are such that $\gidq \geq 0$ and $\gidq \to 15$ as
        $P \to 100$.
    }
    \label{fig:idq-transform}
\end{figure}

$P_{min}$ and $P_{max}$ are chosen to restrict the dynamic range to $\gidq \in \interval{0}{15}$,
to enforce that candidates are only downranked
as well as limit the dynamic range of $\gidq$. Here, we cap $\gidq$
to be 15, and values below the 50th percentile are mapped to $\gidq = 0$.
Higher values of $\gidq$ suggest a higher degree of non-Gaussianity in the data, so restricting
$\gidq$ to be positive implies that candidate events can only be penalized
from its inclusion into GstLAL.
Restricting $\gidq$ to 15 scales the contribution from iDQ so that it is of
a similar size as a confident event seen from GstLAL ($\log \likelihood \approx 15$).
The mapping between percentile and likelihood is shown in figure \ref{fig:idq-transform}.
After computing the transformation for each timespan, all timespans across the analysis
time of interest are combined and stored as a single file to be ingested by GstLAL.

\subsection{Modification of the Ranking Statistic} \label{sec:ranking-stat}


As mentioned in Sec. \ref{sec:intro} the multi-dimensional
likelihood ratio ranking statistic in GstLAL is responsible for ranking candidate
events. The likelihood ratio is defined as

\begin{align} \label{eq:lnL}
	\likelihood &= \frac{ \cprob{\Dh, \Ob, \SNR, \CHISQ, \DPHI, \DT}{\text{signal}} }
			   { \cprob{\Dh, \Ob, \SNR, \CHISQ, \DPHI, \DT}{\text{noise}} },
\end{align}
where each vector of parameters denotes detector-specific quantities.

With the set of detectors LIGO Hanford (H1), LIGO Livingston (L1), and Virgo (V1),
a vector of parameters for $X$ is given by $\vec{X} = \{X_{H1}, X_{L1}, X_{V1}\}$. 
$\vec{D_H}$ represents the horizon distance, and accounts for the detector sensitivity
at the time of the event.  $\Ob$ represents the detectors which were observing
at the time of the detection. $\rho$ is the detected signal-to-noise ratio (SNR), and
$\xi^2$ is the signal-based-veto, which tests the goodness of fit to the template
waveform. Finally, $\Delta t$ and $\Delta \phi$ are the time and phase difference
between two detectors. The full factorization of the likelihood ratio is described
in \cite{Messick:2016aqy}, \cite{Sachdev:2019vvd}, \cite{Hanna:2019ezx}, and
\cite{Cannon:2015gha}.

For single detector candidates, however, signal consistency and coincidence tests
are not
present as with the multi-detector case. In this case, the main term in likelihood ratio
takes on a simpler form, with the $\Delta t$ and $\Delta \phi$ terms not present:

\begin{align} \label{eq:lnL-single}
	\likelihood &= \frac{ \cprob{D_H, O, \rho, \xi^2}{\text{signal}} }
			   { \cprob{D_H, O, \rho, \xi^2}{\text{noise}} }
               \left(\likelihood_{\textrm{penalty}} \widehat{\likelihood}_{\text{iDQ}}\right)^{-1},
\end{align}

Due to the lack of consistency checks compared with coincident triggers, single
detector triggers are more prone to spurious non-Gaussianities and so have two
additional terms to mitigate these effects. The first is an emperically determined
penalty, $\likelihood_{\textrm{penalty}}$,
ensuring that only single detector triggers which can be cleanly distinguished from
the background are considered to be significant. For this analysis,
$\log \likelihood_{\textrm{penalty}} = 10$. The second term,
$\widehat{\likelihood}_{\text{iDQ}}$,
accounts for non-Gaussian noise and is informed directly by iDQ.

For the iDQ term, a $\pm 1$ second window is applied around the
coalescence time, where $\gidq$ is maximized over this window and
applied to single detector candidates. This term downweighs the significance of the
event candidate by the likelihood that there is a noise transient in the 2 second
window.

\section{Results} \label{sec:results}

Here, we focus on the impact of data quality information incorporated into the
GstLAL search for GWTC-2 results.
We will discuss general statements about the impact of iDQ on rejecting glitches
as well as the bulk properties and impact on the pipeline. We provide two specific
times in O3a as concrete examples. The first is
a single-detector gravitational-wave event, GW190419A, found by GstLAL that was vetoed by
a CAT2 flag. The second is a period of time of poor data quality due to a thunderstorm
in Livingston, which was rejected by iDQ as likely being terrestrial in origin. This
same time was also flagged in a data quality flag offline.

\subsection{Glitch Rejection}

\begin{figure}[!ht]
    \centering
    \includegraphics[width=\columnwidth]{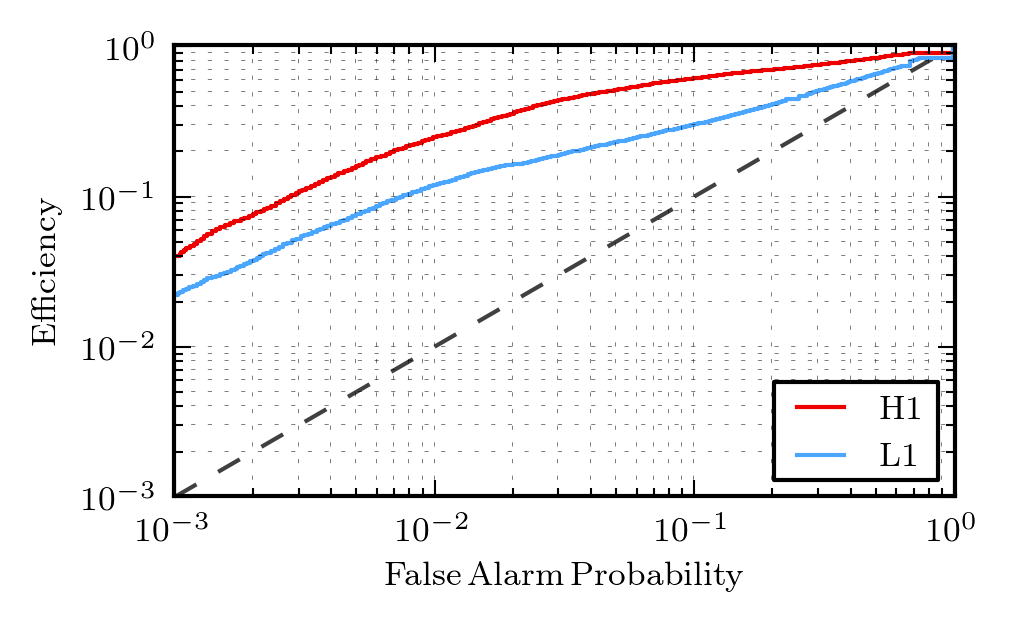}
    \caption{
        Receiver Operating Characteristic (ROC)
        curves for $\gidq$ for H1 and L1.
        The dashed line corresponds
        with a classifier that randomly assigns whether a given time
        corresponds with a terrestrial noise or the absence of it.
        We see that iDQ is better able to identify non-Gaussian noise at H1
        than at L1, likely due to the fact that each detector sees different
        noise sources and has slightly different arrays of auxiliary sensors.
    }
    \label{fig:idq_roc}
\end{figure}

First, we look at the efficiency of iDQ timeseries on glitch rejection. In
Figure \ref{fig:idq_roc}, we see the Receiver Operating Characteristic (ROC)
curves for both H1 and L1, calculated as the fraction of glitches removed
versus False Alarm Probability (FAP), defined as the fraction of time that would
be falsely identified as containing glitches based on the detector's auxiliary
state.
Glitches were determined from times
identified by the Stream-based Noise Acquisition and eXtraction (SNAX) pipeline
\cite{godwin-thesis}
in $h(t)$ with an SNR at or above 8. Similarly, times determined to
be clean were sampled over times where SNAX had no glitches at or above 6. This
criterion was chosen in the same manner as is used for iDQ during low-latency
operation for O3.

It is expected that the performance of using the $\gidq$ timeseries
would be diminished compared to the raw timeseries, due to washing away some
of the knowledge used by iDQ in the calibration stage. However, since the goal
of the transformation
was to take a conservative approach and tackle only the most egregious times of
poor data quality, this is acceptable.
\footnote{Characteristic performance for iDQ during O3 for $\idq$ timeseries
can be seen in \cite{idq-methods}.}
At a $1\%$ false alarm probability, we can
remove $~25\%$ of glitches in H1 and $~12\%$ of glitches in L1 , and at a $0.1\%$
false alarm rate, we can remove $~4\%$ of glitches in H1 and $1-2\%$ in L1.
Note that this also incorporates times in which iDQ was not available due to data
dropouts, and therefore a small fraction of glitches that would otherwise have been
identified may have been missed. Offline iDQ runs would not suffer from such
issues.

\subsection{Search Impact}

\begin{figure}[!ht]
    \centering
    \includegraphics[width=\columnwidth]{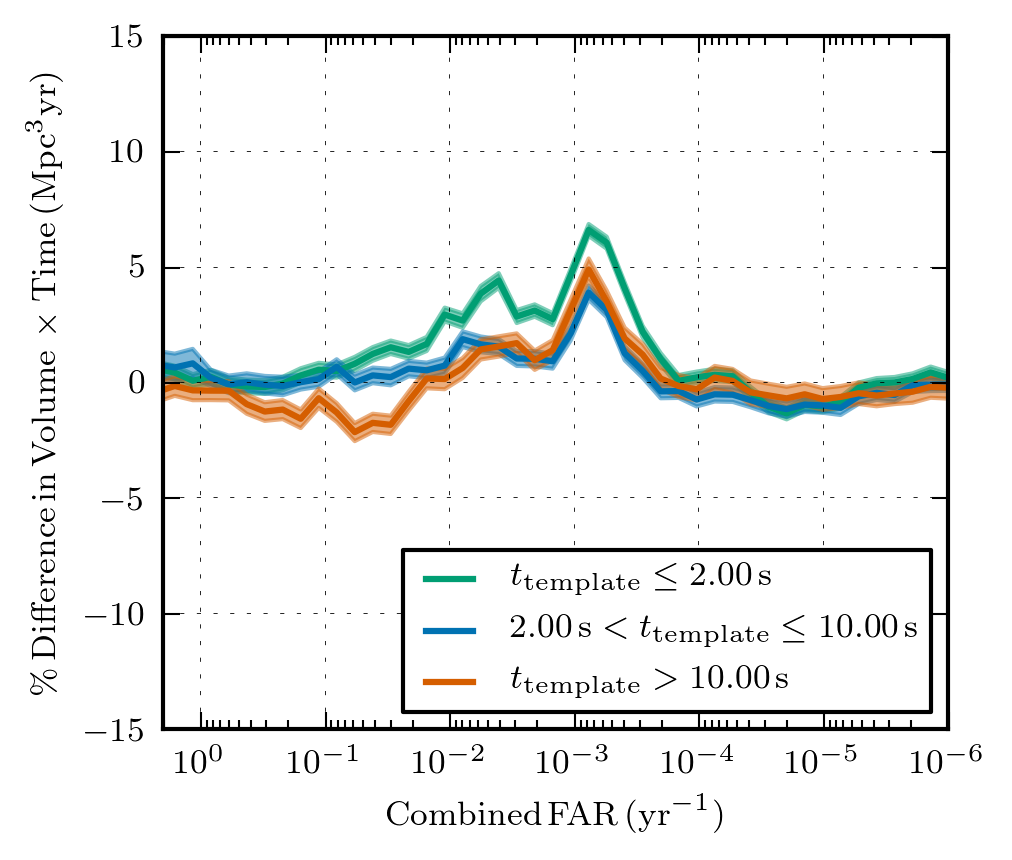}
    \caption{
        Change in search sensitivity with and without
        iDQ information included as a term in the ranking statistic,
        binned by $t_{\mathrm{template}}$ corresponding to the
        template duration of the compact binary waveform.
        The shaded regions represent $90 \%$ credible intervals.
        $t = 2 \,\mathrm{s}$ is the window
        in which $\gidq$ is maximized over
        and applied to single-detector event candidates.
        The high FAR limit
        is capped at the GWTC-2 FAR threshold at 1 per 6 months.
    }
    \label{fig:vt_comparison}
\end{figure}

While the ROC curves demonstrate the recalibrated iDQ timeseries' ability to
identify non-Gaussian noise efficiently, it is not clear that the identified noise
actually limits the search's sensitivity. As such,
we look at the impact in the GstLAL pipeline over GWTC-2 results.
The sensitivity of the search pipeline is typically measured by performing
injections with simulated gravitational-wave signals and measuring the number
of injections that are recovered, providing a measure of the sensitive volume
\cite{TheLIGOScientific:2017lwt}.
This quantity is multiplied by the analysis time to give Volume x Time (VT),
taking into account any removed time from the analysis when applying vetoes.
For an astrophysical population distributed uniformly in co-moving volume and
source-frame time, the expected number of detections is directly proportional
to VT. Any improvement in VT should directly correspond to more detected events.
Figure \ref{fig:vt_comparison} gives the change in search sensitivity
with and without iDQ information incorporated into
the analysis.

Improvements in VT are seen across all template waveforms
in the intermediate FAR region at the few percent level, and appears to be
more significant
for templates fully encompassing the window where $\gidq$ is applied. 
A slight decrease in VT is seen at $\mathrm{FAR} \approx 10^{-5}$ for
shorter duration templates, which
is not well-understood but should not cause recovered
candidates found with high significance to be rejected. A decrease
in VT is also seen for templates longer than $t = 2 \,\mathrm{s}$
at higher FARs which may be due to only considering $\gidq$ immediately
surrounding the merger time.

Looking at the candidate list, a new candidate event is found which was
vetoed previously in a
data quality flag. In addition, using iDQ information has downranked a
potential candidate found previously by GstLAL and associated with a thunderstorm
in LIGO-Livingston. This thunderstorm was flagged as being significant by the
iDQ analysis. It was also added as a data quality flag independently by
the Detector Characterization group for use in offline analyses.

\subsubsection{GW190424A} \label{sec-GW190424A}

This event was found by GstLAL as a single detector BBH in Livingston.
However, the time surrounding this event was vetoed by an
offline CAT2 flag. This veto in particular was created due to periodic glitching from a
camera shutter \cite{camera-shutter-alog}. An accelerometer near the Y-end
test mass witnesses the
source of the glitches and was used to create a veto for offline studies.
This glitching coincided with a gravitational-wave
signal and applying a veto would also veto the gravitational-wave event.

\begin{figure}[!h]
    \centering
    \includegraphics[width=\columnwidth]{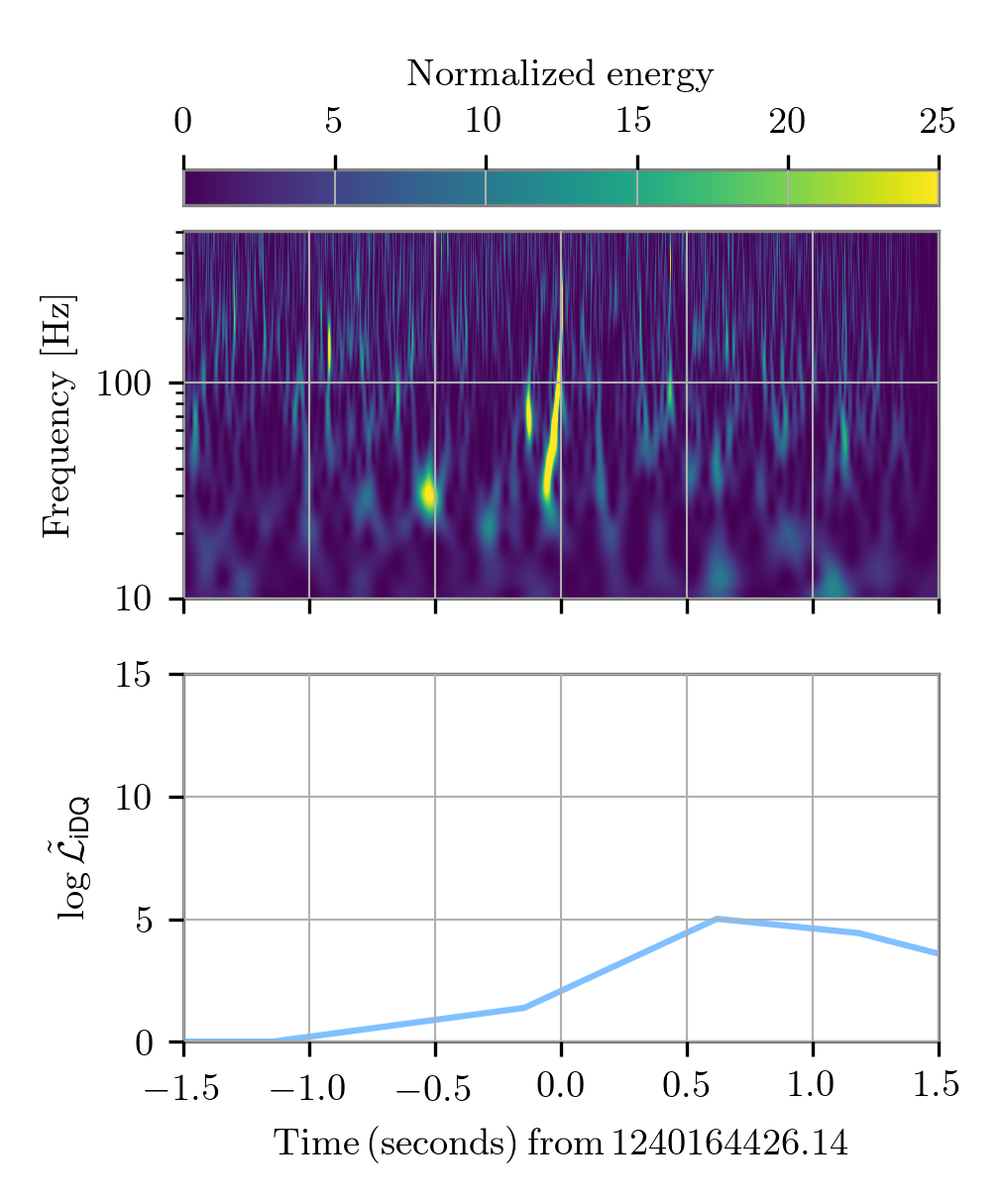}
    \caption{
         Time surrounding GW190424A in LIGO-Livingston,
         shown in a time-frequency spectrogram \cite{Chatterji:2004qg}
         (above) and in the $\gidq$ timeseries (below). A clear inspiral track
         is visible in the time-frequency spectrogram,
         indicating a strong signal candidate. While the $\gidq$
         term in GstLAL's ranking statistic heavily penalizes the timespan surrounding
         this event, other terms in the ranking statistic significantly points to
         the event being signal-like so that the candidate is recovered in an offline
         analysis.
    }
    \label{fig:idq_GW190424A_online}
\end{figure}

In Figure \ref{fig:idq_GW190424A_online}, one can see the inspiral track as
well as excess noise from camera glitching about 0.2 seconds before.
iDQ also has a peak around the
same area and continuing on 1-2 seconds after the merger due
to ringing in response to the initial impulse witnessed by the beamtube accelerometer
\cite{camera-shutter-alog}.
Since information provided by iDQ can only downrank a candidate
event, GW190421A was not vetoed but merely downranked significantly due to the
presence of the glitch. Despite that, the GstLAL analysis found this event as
significant enough at the catalog threshold of 1 per 6 months.


\subsubsection{Thunderstorm Event}

\begin{figure}[!h]
    \centering
    \includegraphics[width=\columnwidth]{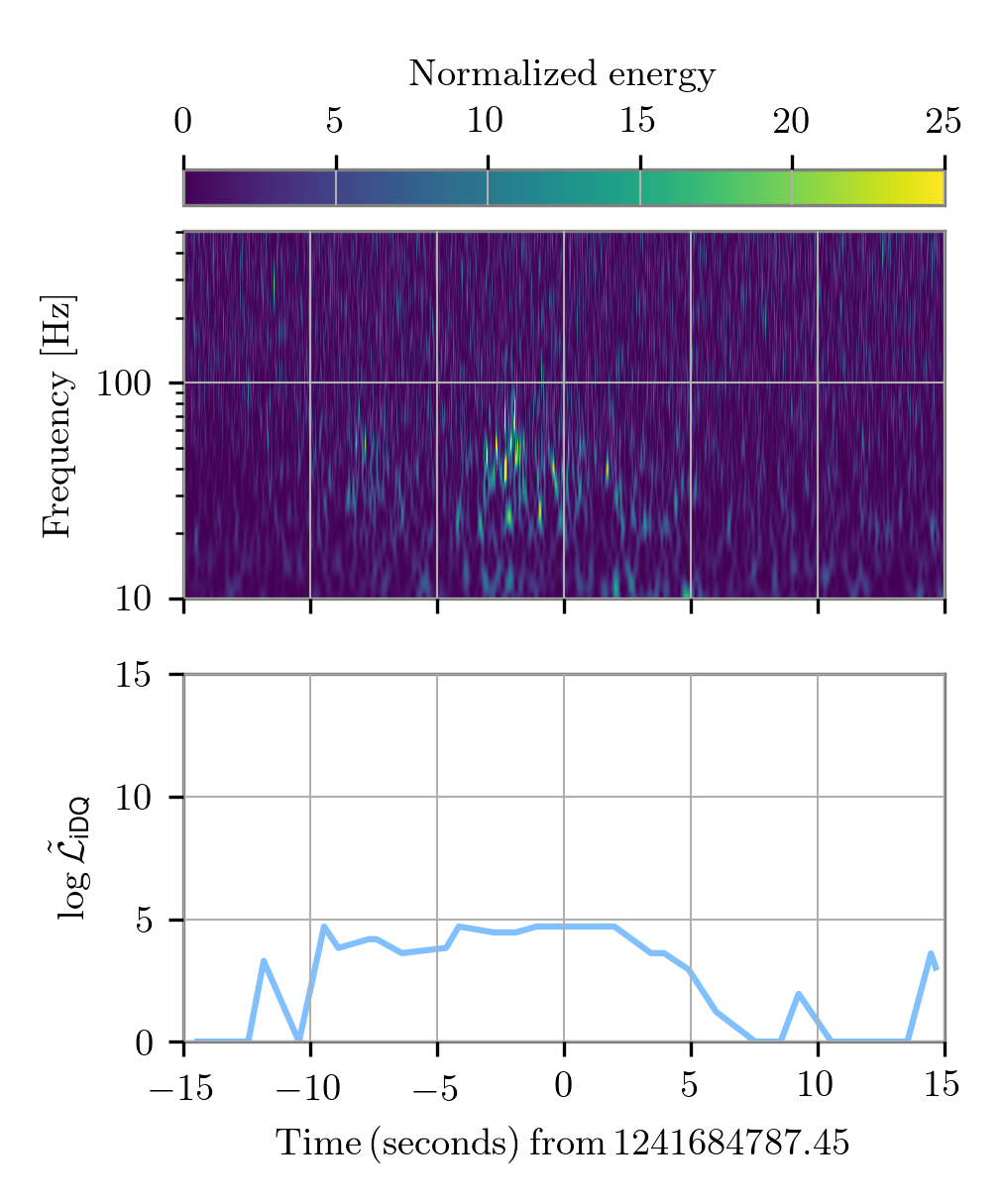}
    \caption{
         Non-Gaussian noise in $h(t)$ caused by a thunderstorm in LIGO-Livingston,
         shown in a time-frequency spectrogram (above) and in the $\gidq$
         timeseries (below). Excess noise is seen in the time-frequency spectrogram
         due to the thunderstorm, corresponding
         to periods of non-Gaussianity identified by iDQ.
    }
    \label{fig:idq_thunder_online}
\end{figure}

Here, we look at a timespan of a few seconds of detector non-Gaussianity
caused by a nearby thunderstorm in LIGO-Livingston during the third observing
run. This time was flagged and captured in a data quality veto offline by the
Detector Characterization group after it was found that the poor data quality in
$h(t)$ in this
stretch of time in L1 was caused by a thunderstorm. It was initially flagged by
GstLAL as a potential single-detector candidate in an offline analysis without using
vetoes or iDQ information. After incorporating iDQ information into the analysis, the
candidate event was heavily downranked and no longer was considered to be a
suitable candidate. In Figure \ref{fig:idq_thunder_online}, we see the iDQ
timeseries for L1 around the time of the thunderstorm event and that iDQ flags
the time as glitchy.

\section{Conclusion} \label{sec:conclusion}

We demonstrated a method of handling transient noise in GstLAL by including
iDQ information as a term in GstLAL's ranking statistic, and saw the impact
on GWTC-2 search catalog results by looking at single-detector gravitational-wave
event, GW190424A and a period of non-Gaussianity caused by a thunderstorm.
A clear benefit of using non-binary data quality information arises from
downranking periods of non-Gaussianities and preventing a scenario where
a gravitational-wave candidate in coincidence with a glitch could be vetoed.
Note that while a gravitational-wave candidate can not be vetoed with this method,
marginal candidates that fall below the threshold for consideration in being a candidate
may still be lost after downranking the event. 
In addition, incorporating iDQ information available in near-real-time
alongside $h(t)$ paves the way for autonomously folding in data quality information
from the detector's auxiliary state for a wider set of glitch classes.

A few aspects could be further improved from the current implementation. First,
since the analysis expects the iDQ timeseries to be available for all times ahead of time,
this implementation currently doesn't work in the low-latency scenario. Moving forward,
it would be ideal to allow the iDQ timeseries to be collected in low-latency during the
filtering stage and fed into the ranking statistic directly while collecting
event candidates. However, the incorporation of iDQ information in low-latency and
subsequent event identification in the presence of glitches may still impact parameter
estimation \cite{Pankow:2018qpo} and sky localization estimates, and isn't addressed
in this work. Allowing iDQ data products to be fed in
asynchronously would also be beneficial for the offline case for the purposes of
reranking candidates, since the filtering stage is costly compared to subsequent
candidate rerankings.

Second, a more comprehensive study of applying iDQ information, including
choosing an optimal window around the candidate event, whether to maximize or
integrate over $\gidq$ across the window, and whether we can choose a
less conservative approach and directly use $\idq$ would be beneficial.

Additionally, any improvements
to iDQ in performing its statistical inference may be reflected directly
as improvements in VT in the GstLAL search. Many of the improvements within iDQ
in O3 were targetted at improving the latency so that data products generated
by iDQ could be available concurrently with calibrated $h(t)$. Moving forward,
being able to provide iDQ a richer set of auxiliary information to perform
its statistical inference such as time-frequency representations may improve
iDQ's ability to detect non-Gaussian noise.
This may also be combined with algorithmic
improvements that would leverage a richer feature set.

Finally, the incorporation of iDQ was limited solely to single detector candidates,
but it would be beneficial to apply this to coincident events as well. One
complication from using $\gidq$ timeseries in coincident events arises
due to not allowing candidates to be upranked from iDQ after the transformation
is applied.
Since on average there was a small contribution coming from each detector,
this caused coincident events to be downranked a bit too strongly since we were
effectively applying a penalty multiple times.
This issue may end up being much less of a concern if iDQ
information is also allowed to uprank during clean times.

We demonstrated an approach of folding in data quality
information via iDQ into the GstLAL search without losing livetime
and mitigating the possibility of vetoing gravitational-wave candidates.
Since iDQ is available alongside $h(t)$, this gives a clear path to allow the
inclusion of data quality information leveraging the detector's auxiliary state
in low latency. This also moves us towards understanding how to optimally
apply non-binary data quality information in a gravitational-wave search,
which will be increasingly important as the detector network sensitivity
increases in future observing runs.

\acknowledgements
This work is supported by the National Science Foundation (NSF) through OAC-1841480,
OAC-1642391, and PHY-1454389.
The authors are grateful for computational resources provided by the
LIGO Laboratory and supported by NSF grants PHY-0757058 and PHY-0823459.
Computations for this research were also performed on the Pennsylvania State
University’s Institute for Computational and Data Sciences Advanced
CyberInfrastructure (ICDS-ACI).
We thank Jolien Creighton, Keith Riles and Gabriele Vajente for their help
in the review of the inclusion of iDQ information into GstLAL.
R.E. is supported at the University of Chicago by the Kavli Institute for
Cosmological Physics through an endowment from the Kavli Foundation and its founder
Fred Kavli.
J.C. is supported by NSF grant PHY-1912649. D.M. is supported by NSF grant PHY 14-54389,
ACI grant 16-42391, and OAC grant 18-41480.

\appendix \label{sec:appendix}

\bibliography{references}

\end{document}